\documentclass[epj]{mysvjour_fermilab}
\usepackage{graphics}
\def\met{\mbox{${\hbox{$E$\kern-0.6em\lower-.1ex\hbox{/}}}_{T}$}}
\begin{document}
\title{Measurements of Jet and Multijet Cross Sections with the CDF Detector}
\author{Matthias T\"onnesmann\inst{1,2} (Representing the CDF Collaboration)}
\institute{Max-Planck-Institut f\"ur Physik, F\"ohringer Ring 6, 80805
M\"unchen, Germany \and Department of Physics and Astronomy, Michigan State
University, East Lansing, MI 48824, U.S.A.}
\date{Received: date / Revised version: date}
\abstract{
Recent measurements of jet and multijet production cross sections from
$p\bar{p}$ collisions recorded with the Collider Detector at Fermilab (CDF) are
summarized. First Run~II results of the inclusive one jet cross section at
$\sqrt{s} = 1.96$\,TeV as well as prospects for future extensions of this
measurement are presented. We also studied the properties of three-jet events in
Run~Ib data at $\sqrt{s} = 1.8$\,TeV. All results are compared to predictions of
Quantum Chromodynamics at next-to-leading order perturbation theory.
}
\authorrunning{Matthias T\"onnesmann}
\maketitle

\section{Inclusive one jet cross section}
\label{sec:OneJetCrossSection}

One of the most important goals of QCD measurements at hadron colliders is the
extraction of the input parameters of the theory, the strong coupling constant
$\alpha_S$ and the parton distribution functions (p.d.f.). The production of
hadronic jets at the Tevatron also probes the highest momentum transfer region
currently accessible and thus is potentially sensitive to a wide variety of new
physics.

CDF Run~I data~\cite{ref:Run1JetCrossSection} exhibited an excess in the
inclusive jet cross section at high $E_T$ when compared to QCD predictions at
next-to-leading order (NLO) using then-current parton distribution functions.
This excess can be explained by an underestimated gluon content of the proton at
high momentum fraction $x$. Indeed, the gluon distribution is not well
constrained at high~$x$ and has increased in recent p.d.f.\
fits~\cite{ref:CTEQ6}, leading to better agreement with both the CDF and D\O\
inclusive jet cross section measurements.

In Run~II the measurement of jet production and the sensitivity to new physics
will profit from the large integrated luminosity and the higher cross section,
which is associated with the increase in the center-of-mass energy from 1.8\,TeV
to 1.96\,TeV.

\subsection{Status of Run~II measurement}
\label{subsec:Run2JetCrossSection}

The results presented here are based on data recorded from February 2002 through
January 2003 corresponding to an integrated luminosity of 85\,pb$^{-1}$. We have
utilized the same techniques used in the previous CDF Run~I inclusive jet
analysis~\cite{ref:Run1JetCrossSection}. In particular, we apply the Run~I cone
algorithm ({\sc Jetclu}~\cite{ref:JetClu}, $R_\mathrm{cone} = 0.7$) to
reconstruct jets in the central pseudorapidity region ($0.1 < |\eta| < 0.7$).
Events were collected using 4 different $E_T$ trigger thresholds with
appropriate prescale factors. To reduce background from cosmic rays, accelerator
losses, and detector noise, cuts on the missing $E_T$ significance,
$\widetilde{\met} = \met/\sqrt{\sum E_T}$, are applied. A good energy
measurement of jets is ensured by requiring the event vertex to be within 60\,cm
of the center of the detector along the beam direction.

The measured jet energies are corrected for experimental effects stemming from
non-uniformities of the calorimeter response, multiple interactions, calorimeter
non-lin\-e\-ar\-i\-ty, and energy due to the underlying event. Since we
currently rely on the absolute energy corrections determined in Run~I, the jet
energy scale has been set to that of Run~I, thereby introducing a systematic
uncertainty of 5\,\%, which is the dominant experimental systematic error.
Further understanding of the energy scale will reduce this uncertainty.

\begin{figure*}[t]
\resizebox{0.489\textwidth}{!}{\includegraphics{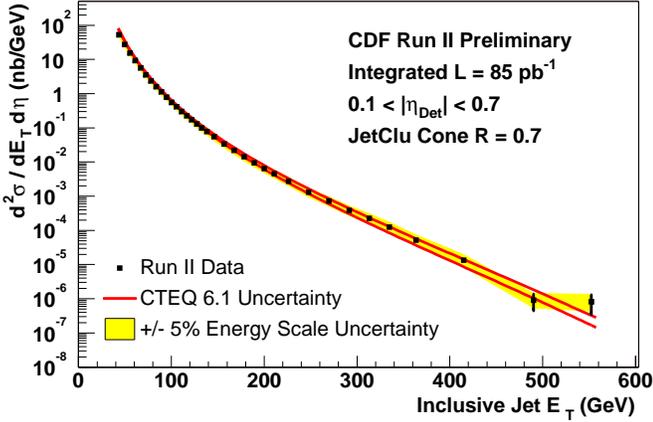}}
\hfill
\resizebox{0.489\textwidth}{!}{\includegraphics{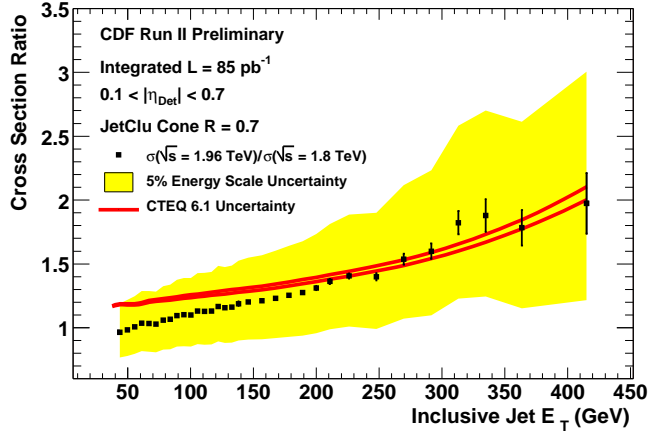}}
\caption{({\it Left}\/) Comparison of the measured inclusive jet cross section
from Run~II data (points) to QCD predictions at NLO using CTEQ\,6.1 p.d.f.\
(curves). ({\it Right}\/) Run~II/Run~I cross section ratio compared to the QCD
prediction at NLO using CTEQ\,6.1 p.d.f.}
\label{fig:1}
\end{figure*}

The unsmeared jet cross section is shown in Fig.~\ref{fig:1} ({\it left}\/).
It is compared to a QCD prediction at NLO, which reproduces the distribution of
the data well over 8 orders of magnitude. The theoretical prediction was
calculated using the EKS program~\cite{ref:EKS} and the CTEQ\,6.1 p.d.f.\
set~\cite{ref:CTEQ61}. The renormalization and factorization scales were set to
$E_T/2$. The CTEQ\,6.1 set of p.d.f.\ has available complete error information,
which makes it possible to calculate the p.d.f.\ errors on the Run~II jet cross
section predictions, indicated as the curves in Fig.~\ref{fig:1}. The dominant
p.d.f.\ uncertainty comes from the gluon density at high~$x$, which is the least
well constrained parameter of the CTEQ\,6.1 p.d.f.\ set. The full Run~II
dataset will help to reduce this uncertainty (see Sect.~\ref{subsec:Prospects}).

The effect of the higher jet cross section in Run~II is especially prominent at
the high $E_T$ frontier, where two new bins were added. With a data sample
similar in size to that obtained in Run~Ib, we are thus already able to extend
the $E_T$ range covered by the Run~I measurements by almost 150\,GeV. The
Run~II/Run~I cross section ratio, together with a QCD prediction at NLO, is
shown in Fig.~\ref{fig:1} ({\it right}\/). The ratio is seen to be lower than
expected at low $E_T$, but we find good agreement within the experimental and
theoretical uncertainties.

\subsection{Prospects for future extensions}
\label{subsec:Prospects}

A powerful way to understand the nature of a potential excess in the jet cross
section at high $E_T$ is the extension of the analysis described above into the
forward region of the detector. The new CDF endplug calorimeters, which cover
the pseudorapidity range $1.1 < |\eta| < 3.6$, will permit such a measurement.
Forward jet measurements are not expected to have any contribution from new
physics because the maximum reachable $E_T$ is limited to, e.g., about 200\,GeV
for $2.1 < |\eta| < 2.8$. On the other hand the sensitivity to the gluon
distribution in the proton is similar to that of central jet measurements. The
gluon distribution at high~$x$ can thus be further constrained, which will in
turn increase the sensitivity to new physics in the high $E_T$ (high mass)
region of the central one jet (di-jet) cross section.

Another improvement in jet measurements can be attained by the use of other jet
reconstruction algorithms. CDF has so far relied on its cone algorithm
{\sc Jetclu}~\cite{ref:JetClu} to search for jets, define jet observables and
measure jet cross sections. During the past few years different theoretical
problems of cone algorithms were pointed out~\cite{ref:Run2Workshop}, namely the
infrared and collinear sensitivity of the observables, e.g.\ cross sections, and
the difficulty to match the experimental algorithms with those employed in
theoretical calculations. Besides improved cone algorithms, the longitudinally
invariant $k_T$ clustering algorithm~\cite{ref:KtClus} will be an important tool
because of its built-in infrared and collinear insensitivity and its direct
applicability at the parton and at the detector level.

\begin{figure}[b]
\resizebox{0.489\textwidth}{!}{\includegraphics{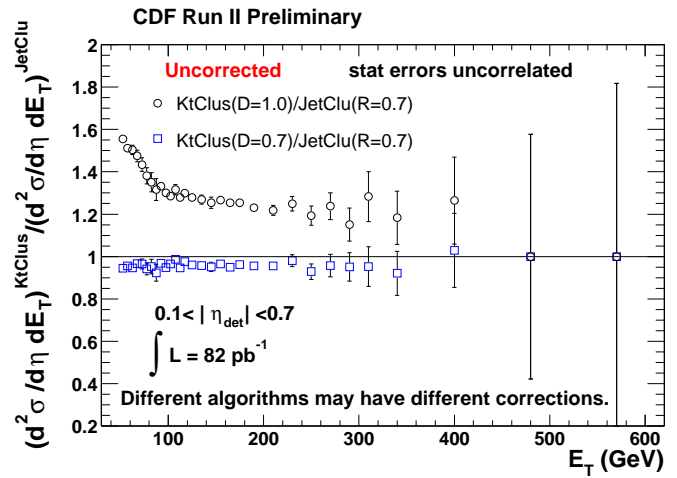}}
\caption{Ratio of raw jet cross sections using the $k_T$ clustering algorithm
($D = 0.7, 1.0$) and the {\sc Jetclu} cone algorithm ($R_\mathrm{cone} = 0.7$)
from Run~II data.}
\label{fig:2}
\end{figure}

Fig.~\ref{fig:2} shows the ratio of raw (uncorrected) inclusive one jet cross
sections using the $k_T$ clustering algorithm with the angular jet separation
parameter $D$ set to 0.7 and 1.0 and the {\sc Jetclu} algorithm
($R_\mathrm{cone} = 0.7$).
Events were selected as described in Sect.~\ref{subsec:Run2JetCrossSection}. For
$D = 0.7$ the uncorrected $k_T$ cross section is about 5\,\% lower than the
{\sc Jetclu} cross section, while $D = 1.0$ produces bigger jets with larger
$E_T$, which directly translates into a 20\,\% increase in the cross section.
Furthermore we observe an increase of the ratio at low $E_T$, which is
qualitatively similar to the low $E_T$ behavior of the ratio of fully corrected
and unsmeared cross sections measured by D\O\ using $k_T$ jets ($D = 1.0$) in
the Run~I data sample~\cite{ref:DZeroKt}. It is important to note, however, that
different jet algorithms may have different energy corrections. In particular,
the correction for the underlying event will be larger for $k_T$ jets with
$D = 1.0$ than for $D = 0.7$. Quantitative comparisons should therefore be
carried out only after correcting the jet energies and unsmearing the cross
section distribution.

\begin{figure*}[t]
\resizebox{0.442\textwidth}{!}{\includegraphics{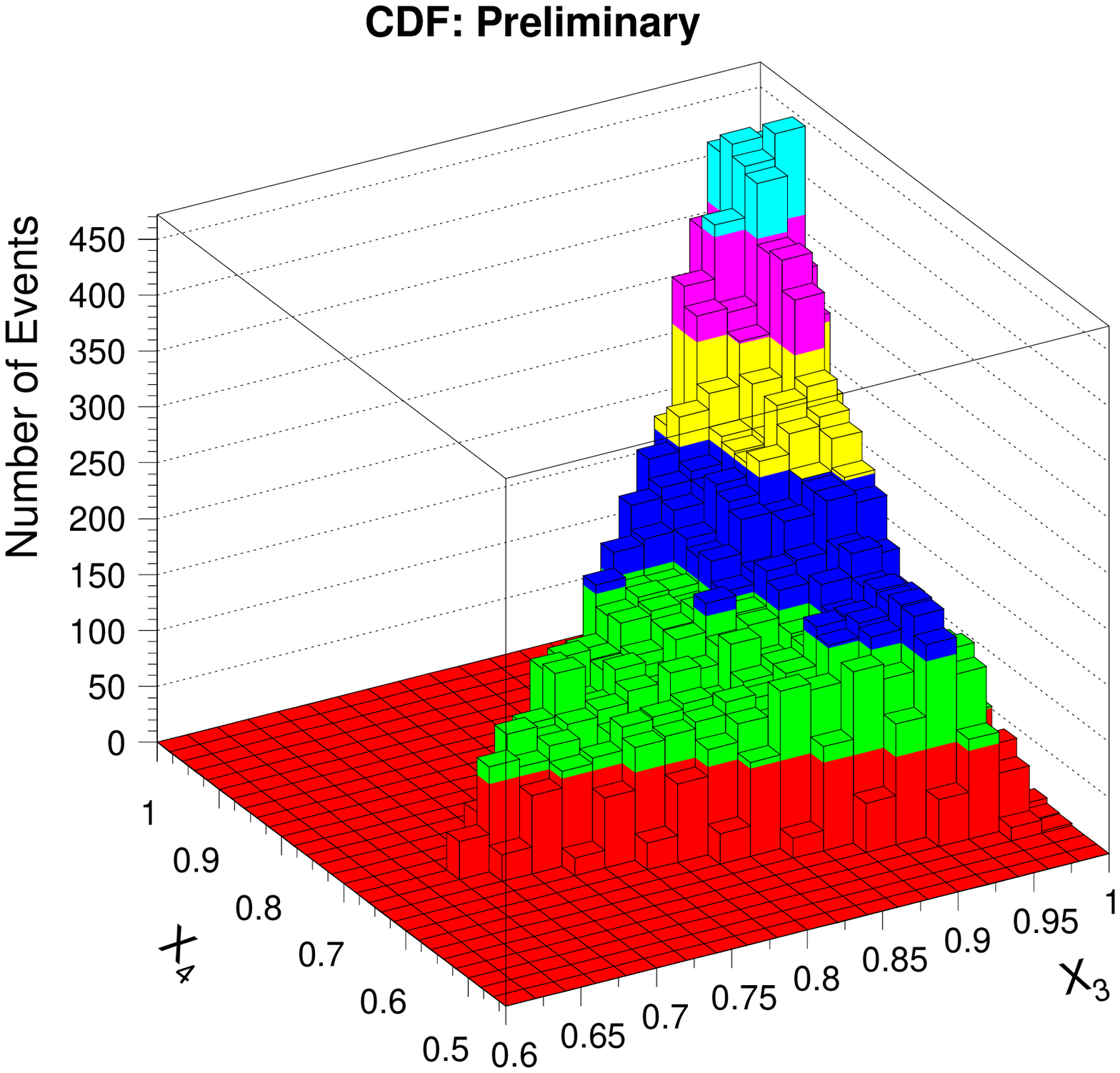}}
\hfill
\resizebox{0.503\textwidth}{!}{\includegraphics{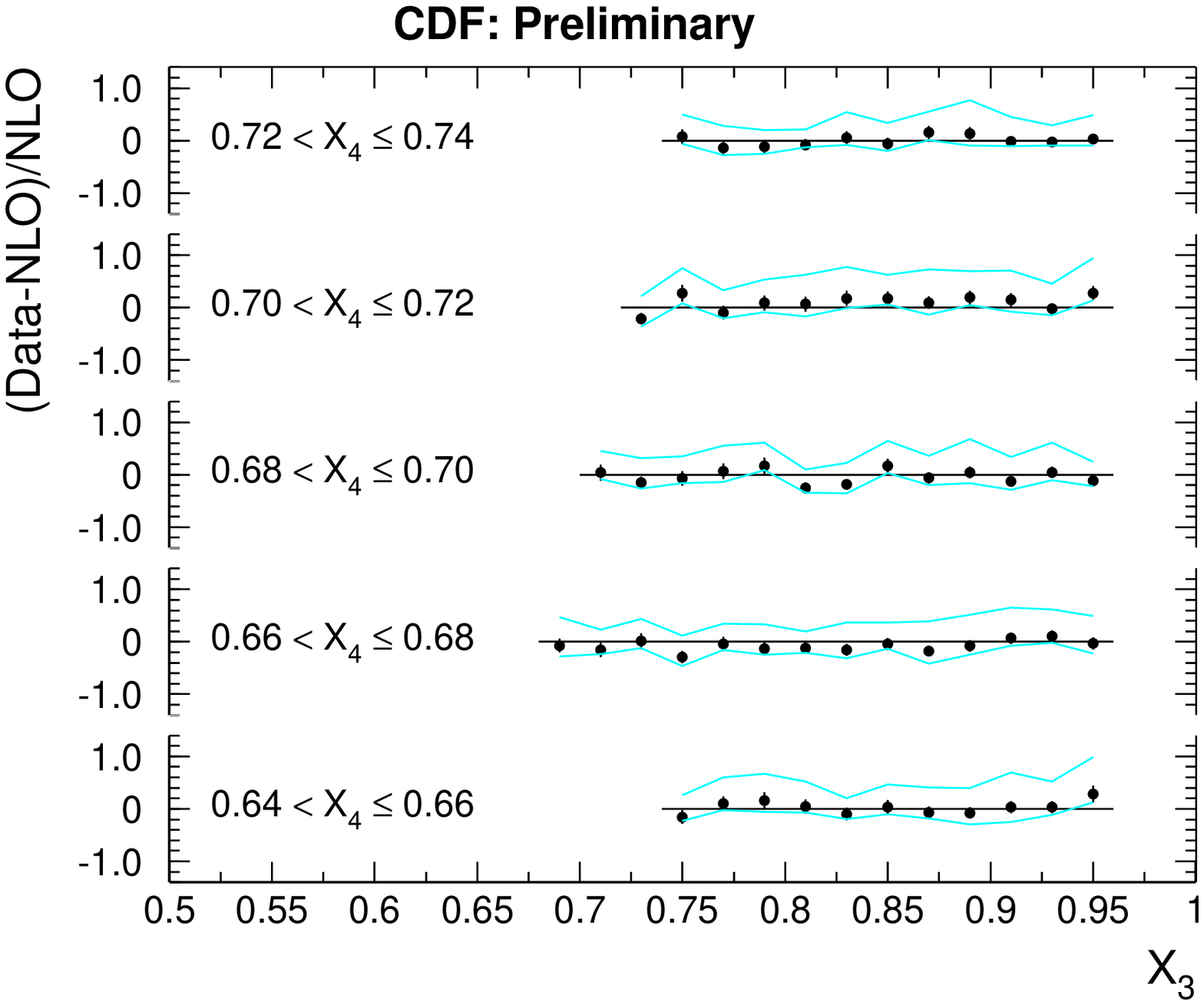}}
\caption{({\it Left}\/) Measured distribution of the 3-jet events in the
$X_3$--$X_4$ plane from Run~Ib data. ({\it Right}\/) Ratio
$(\mathrm{Data} - \mathrm{NLO})/\mathrm{NLO}$ for the differential cross section
as a function of $X_3$ in the region $0.64 < X_4 \leq 0.74$. The band between
the two curves represents the experimental systematic uncertainties.}
\label{fig:3}
\end{figure*}

The long term goal is to measure the inclusive jet cross section using the
improved Run~II cone algorithm and the $k_T$ clustering algorithm, and to extend
the measurements to the forward region.

\section{Three-jet cross section}
\label{sec:ThreeJetCrossSection}

With the availablitity of QCD predictions at NLO for the production of 3-jet
events at hadron colliders~\cite{ref:NLO3Jet} new possibilities for precision
tests of QCD have opened up, among them the measurement of $\alpha_S$ from the
ratio of 3-jet and 2-jet production rates or from event shapes. A different
approach is the analysis of the topology of 3-jet final states using Dalitz
variables, which will be presented in the following.

We analyzed 86\,pb$^{-1}$ of Run~Ib data. Jets are reconstructed using the
{\sc Jetclu} algorithm with $R_\mathrm{cone} = 0.7$. 3-jet events are selected
by requiring at least 3 jets with $E_T \geq 20$\,GeV and $|\eta| < 2.0$,
$\sum E_T(\mbox{3 jets}) > 320$\,GeV, and a separation of $\Delta R > 1.0$ in
the eta--phi plane between the jets. The events are boosted into the 3-jet rest
frame, and the 3 leading jets are numbered such that $E_3 > E_4 > E_5$. The
3-jet mass $m_\mathrm{3\mbox{-}jet}$ is calculated, together with the Dalitz
variables $X_i = 2E_i/m_\mathrm{3\mbox{-}jet}$, $X_3 + X_4 + X_5 = 2$.

Fig.~\ref{fig:3} ({\it left}\/) shows the measured distribution of the 3-jet
events in the $X_3$--$X_4$ plane. The topologies are dominated by configurations
containing a soft third jet.

The differential cross section as a function of $X_3$, measured in different
bins of $X_4$, was compared to QCD calculations at NLO using the CTEQ\,4M p.d.f.
Fig.~\ref{fig:3} ({\it right}\/) shows the relative difference between data and
theory in the region $0.64 < X_4 \leq 0.74$. Reasonable agreement was observed
in the whole $X_3$--$X_4$ plane. The experimental systematic uncertainties are
dominated by the jet energy scale.

The total 3-jet production cross section, integrated over the $X_3$--$X_4$ plane
with $X_3 < 0.98$, yields
$\sigma^\mathrm{3\mbox{-}jet} = 456 \pm 2\,\mbox{(stat.)}
^{+202}_{-68}\,\mbox{(syst.)}$\,pb$^{-1}$, consistent with the NLO prediction
$\sigma^\mathrm{3\mbox{-}jet}_\mathrm{NLO} = 482 \pm 2\,\mbox{(stat.)}
^{+31}_{-72}\,\mbox{(theo.)}$\,pb$^{-1}$. The theoretical uncertainty is due to
the arbitrary choice of the renormalization and factorization scales, and was
calculated by varying the scales by factors of 0.5 and 2.

The NLO predictions have also been calculated using different members of the
CTEQ\,4A p.d.f.\ family~\cite{ref:CTEQ4}, which differ from CTEQ\,4M in the
value of $\alpha_S$. However, an extraction of $\alpha_S$ from a $\chi^2$
analysis is not possible due to lack of sensitivity to $\alpha_S$ within the
large uncertainties.

\section*{Acknoledgments}
I would like to thank the European Commission and the European Physical Society
for financial support.

\end{document}